\documentclass[prb,preprint]{revtex4-1} 

\usepackage{amsmath}  
\usepackage{amsfonts} 
\usepackage{graphicx} 

\begin{document}


\title{To the abilities of optical triode: optical triode-based RF generator.}

\author{Gleb G. Kozlov}
\email{gkozlov@photonics.phys.spbu.ru} 
\affiliation{Spin-Optics laboratory, St.~Petersburg State University, 198504 St.~Petersburg} 


\author{Valentin G. Davydov}
\email{v.davydov@spbu.ru}
\affiliation{Photonics dept., St.~Petersburg State University, 198504 St.~Petersburg}



\begin{abstract}
The optical analog of vacuum triode with electron flow being replaced by photon flow --- 
optical triode (OT) --- is considered. Distinctions of such a device with respect to vacuum and 
semiconductor triodes are discussed. As an illustration and example of possible application of OT 
the design of RF generator without conventional active elements is experimentally demonstrated.
The amplification and feedback are realised in optical channel by means of Pockels cell and vacuum 
photodetector. The application of suggested device as a sensitive photo receiver is discussed. 
\end{abstract}

\maketitle 

\section*{Introduction} 
 
The fantastic development of radio technology and the use of electronic devices 
in almost all areas of human activity began with the invention by Lee de Forest 
in 1906 of a high-speed electric signal amplifier --- a vacuum triode \cite{f1,f2,f3,f4,f5}.
Despite the fact that currently vacuum triodes have mostly given way to solid state 
semiconductor amplifying elements, both of them have many common features, and many 
circuitry solutions found in the era of electro-vacuum amplifying devices still remain 
relevant. The simple design of a vacuum triode clearly illustrates the very principle 
of the amplifying element as a device in which relatively weak energy flow controls 
a stronger one. In the Lee de Forest vacuum triode, the electron flux arising due to 
thermionic emission from the cathode and the accelerating electric field of the anode 
is modulated by electric field of an electron-transparent grid placed between the anode 
and cathode. Changes in the grid potential leads to change (up to complete blocking) 
of the electron current from the cathode to the anode, and the electric power in the 
anode circuit can significantly exceed the power in the grid circuit (power gain).
       
In the proposed article, it is shown that a similar structure of the amplifying element 
can be completely reproduced in a device in which the electron beam is replaced by a light 
(photon) beam. The suggested device --- an optical triode --- is implemented with simple 
laboratory means and can serve to explain the principles of amplification and demonstration 
of elementary circuitry solutions in school and student classrooms.
        
The article is organized as follows. The first section describes the design of the optical 
triode, introduces the definition of its parameters, and compares the optical triode with 
conventional amplifying elements. The second section describes sample design of an 
optical-triode-based RF oscillator and provides its theoretical analysis. The third 
section describes the experimental implementation of the RF generator and discusses 
its usage as an active photodetector. The results of the paper are briefly
summarized in Conclusion.
    
   \section{The optical triode}
   Consider the electro-optical modulator (Pockels cell with associated polarization elements) and 
   the photodetector (semiconductor or vacuum photodiode) sequentially arranged in a light beam. 
   The variation of voltage $U$ on the electro-optical modulator will cause a change in transmitted
   light intensity and hence the photocurrent $I$. If $ U_ {\lambda / 2} $ is the  electro-optical 
   modulator half-wave voltage, then the following estimate holds
   \begin{equation}
   {dU\over dI}\equiv {1\over S}\sim {U_{\lambda /2}\over I_0}
   \label{1}
   \end{equation}
   where the photocurrent of ``open" modulator $ I _0 $ can be derived from light beam intensity $ P $ 
   and quantum efficiency of the photodetector $ \eta $ by the relation:
   $$I_0={\eta P e \over \hbar \omega},$$
   where $e$ is electron charge and $\hbar \omega$ --- photon energy.
   Described sequence of electro-optical elements (which we will call  
   {\it optical triode}) works similar to electro-vacuum triode:
   light source plays role of the cathode,  electro-optical modulator --- 
   grid, and photodetector --- anode. By analogy with the vacuum triode, 
   we consider transconductance $ S $ of the optical triode. Due to the 
   negligible electrical conductance of the Pockels cell, the possibility 
   of power amplification by an optical triode is quite obvious.
   
   Let us estimate the possible transconductance of optical triode. 
   Take a semiconductor laser with  output power $ \sim $ 1.5 W  as a light 
   source, and a semiconductor photodiode with quantum efficiency  
   $ \eta \lesssim 1 $ as a photodetector. Then current $I_0$ is $ \sim 0.5 $ A. 
   To obtain a low half-wave voltage, the crystal in Pockels cell should be 
   made as thin (in the direction of electric field) as possible. 
   It is known that DKDP crystal thickness in off-the-shelf ML-103 cell is  
   $ \sim 3 $~mm, the crystal length is $ \sim 50 $~mm and a half-wave 
   voltage is $ \sim 200 $ V for visible light. It seems to be feasible 
   to reduce the crystal thickness by 100 times (to 30~$\mu$m) at the cost 
   of length reduction by~10~times (to 5~mm). Then transconductance of 
   the optical triode is expected to be $ S \sim 250 $~ma/V, 
   which is an order of magnitude greater than that of vacuum 
   triodes with similar power consumption. 
   
   The input impedance of optical triode is expected to be greater than that of 
   insulated gate field effect transistor (MOSFET) of comparable transconductance,
   mainly due to lower capacitance of the Pockels cell.
   
   Transconductance $S$ of the optical triode can be linearly controlled by 
   changing power of the light beam, e.~g.\ by an additional optical modulator.
   Therefore one can create voltage controlled amplifiers. 
   
   Yet another valuable advantage 
   of optical triode is its inherent galvanic isolation between input and output. 
   Passthrough capacitance can be almost arbitrarely reduced by increasing the 
   distance of light travel between optical modulator and photoreciver.
   
   \begin{figure}[bhtp]
    \centering
    \includegraphics[scale=0.666666]{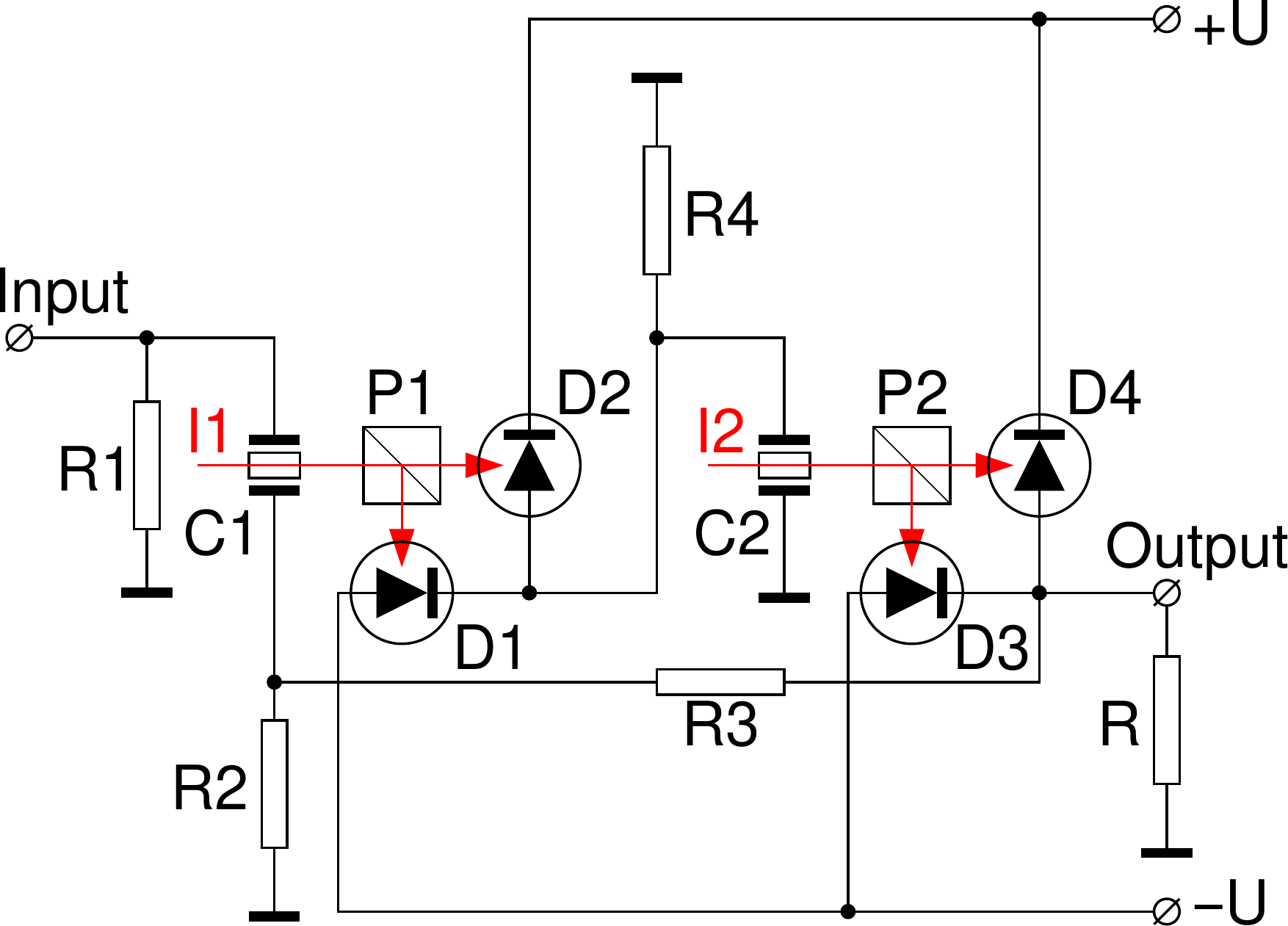}
    \caption{Two-cascade electro-optical amplifier}
    \label{fig1} 
   \end{figure}

   Within the same general principle of electro-optical active element its design
   can be varied depending on the application. As an example, two cascade linear 
   amplifier is presented in Fig.~\ref{fig1}. The polarisation beam splitters 
   P1 and P2 are oriented so that when the input voltages of Pockels cells 
   C1 and C2 are zero, photocurrents of D1, D2 are equal to each other (the
   same for D3, D4). Therefore net currents in load resistors R4 and R are zero (both 
   pairs of photodiodes are balanced).
   If input voltage of Pockels cells becomes nonzero then intensities of horizontal 
   and vertical (as depicted in Fig.~\ref{fig1}) light beams change in opposite directions 
   giving rise to nonzero currents in load resistors R4 and R.
   If most of voltage amplification is provided by the first cascade, then intensity 
   $I_1$ can be chosen relatively small. Required amplification coefficient $K$ can be 
   obtained by choosing the value of resistor $R_4$ sufficiently large. For example, 
   if $S_1=25$ ma/V ($I_1\sim 150$ mW) and $R_4=100$ k$\Omega$, we have $K_1\sim S_1R_4=2500$. 
   To obtain large enough power in the load resistor $R$ the intensity $I_2$ should 
   be sufficient to provide the required current of the second cascade photodetectors. 
   For  $I_2=0.25$~A, $S_2=100$~mA/V,  $\pm U=15$~V and $R=50$~$\Omega$
   the amplification coefficient of the second cascade is
   $K_2\sim S_2 R\sim 5$, and the maximum output power is $P_\text{out}=I_2^2R\sim 3$ W.
   The feedback circuit R3, R2 is used to linearise the amplifier's amplitude response. 
   Resulting amplification coefficient can be approximated as $$K={(R3+R2)\over R2}$$ 
   provided that $K\ll K_1K_2$. 
    Depending on the feedback loop coefficient $K_1K_2/K$,
   total optical length (and hence time delay) from Pockels cells to photodetectors and
   capacitances of Pockels cells and photodiodes, an amplifier with feedback could become 
   unstable. To maintain stability, one can implement the Boucherot cell in parallel to 
   (or even instead of) R4.
    
   In conclusion let us derive the input noise voltage in a frequency band $\Delta\nu$ 
   of the cascade with optical triode from the shot noise of the photocurrent. Calculation 
   gives $$\sqrt{\langle \delta U^2\rangle}=U_{\lambda /2}\sqrt{\hbar\omega \Delta\nu\over P\eta}$$. 
   The estimation of noise voltage in unit frequency band for the amplifier duscussed above
   gives the value $\sim $ 2 nV$/\sqrt{\mathrm Hz}$, which is comparable with best devices
   built upon bipolar transistors.
   
   \section{The RF oscillator on optical triode. Quantitative consideration.}
   
   \begin{figure}[bhtp]
   \centering
   \includegraphics[scale=0.666666]{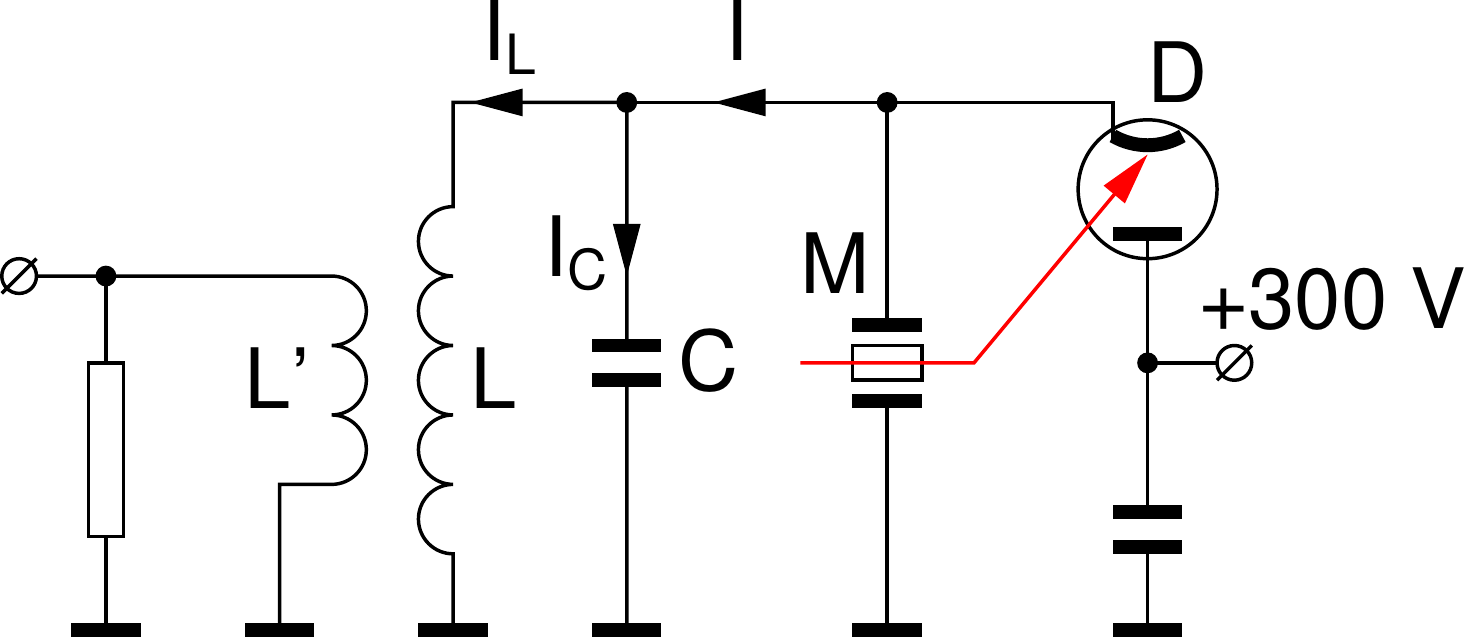}
   \caption{LC oscillator based on optical triode. Polarization optics and parasitic
   impedances are not shown.}
   \label{fig2}
   \end{figure}
   
   Authors were limited in facilities to implement in hardware full-scale amplifier 
   presented in the previous section. Nevertheless the possibility of obtaining the 
   optical-triode-based amplification was demonstrated in a simple experiment described 
   below. Let us consider the optical-triode-based RF generator (Fig.~\ref{fig2}). 
   The light beam after passing through the electro-optical modulator~M connected 
   to the LC tank hits the photocathode of vacuum photocell~D which is also connected  
   to the same tank thus forming the feedback.
   
   The DC components of currents and voltages in the schematics Fig.~\ref{fig2} 
   affect only power consumption and will be omitted from further analysis. 
   Hereinafter only AC components of all voltages and currents are considered.
   Then current $I$ of the photodetector can be expressed in terms of
   voltage $U$ on the electro optical modulator as $I=SU-S_3U^3$, 
   were $S$ is defined by Eq.~(\ref{1}) and the cubic term represents general
   nonlinearity which is necessary for obtaining definite amplitude of steady 
   state oscillations irrespective to the initial conditions. Denoting voltage 
   on the capacitor as $U_C$, current of the photodetector as $I$, current in 
   the coil as $I_L$, active resistance of the coil as $R$ (not shown in the 
   Fig.~\ref{fig2}), one can write the following system of equations for 
   these quantities:

   \begin{align}
    I&=SU_C-S_3U_C^3,
    \label{2}\\
    I&=C{dU_C\over dt}+I_L,
    \label{3}\\
    U_C&=L{dI_L\over dt}+RI_L.
   \label{4}
   \end{align}

   By linearization of this system (i.~e.\ ommiting the term $U_C^3$) 
   one can analyse it's stability and obtain the following necessary condition of 
   unstability with respect to oscillations in the system:
   \begin{equation}
   {LS\over RC}={Z^2S\over R}=QZS>1.
   \label{5}
   \end{equation}
   Here $Z=\sqrt{L/C}$ is the characteristic impedance and $Q=Z/R$ is the Q-factor 
   of the tank circuit. The estimation by this formula shows the possibility 
   (not very strong) of generation.
   
   The exact analytical solution of nonlinear system Eqs.~(\ref{2}--\ref{4}) is 
   not possible. Therefore, to obtain an amplitude of the steady 
   state oscillations we have to use some {\em ad hoc} assumptions\cite{fn}, namely 
   i) the solution of the considered system at sufficiently large time 
   is a periodic oscillations of constant amplitude, and
   ii) these oscillations are nearly harmonic with a frequency~$\omega$ close 
   to the resonant frequency $\omega_0=1/\sqrt{LC}$ of the tank circuit. 
   Also we will consider the tank circuit Q-factor to be large: $Q\gg 1$. 
   Solving (\ref{4}) with respect to $I_L$ we obtain
    \begin{equation}
    I_L(t)=I_L(0)\exp \left(-{t\over \tau}\right)+\int\limits_0^t\!\exp \left({t'-t\over \tau}\right)\,
    {U_C(t')\over L}dt,'
    \label{6}
    \end{equation}
   where $\tau\equiv L/R$. In accordance with the above assumptions we set 
    \begin{equation}
    U_C(t)=A\cos(\omega t),
    \label{7}
    \end{equation}
   where $A$ denotes amplitude of oscillations and $\omega\approx \omega_0$. 
   Then from Eq.~(\ref{6}) one can obtain 
   the following equation
   for $I_L(t)$ at $t\gg\tau$:
   \begin{equation}
   I_L(t)=A\left({\sin (\omega t)\over \omega L}+{\cos (\omega t)\over \omega^2 \tau L}\right).
   \label{8}
   \end{equation}
   
   Here we neglect terms of the order $\sim 1/\omega^2\tau^2\approx 1/Q^2\ll 1$. 
   If considered circuit reached the regime when the assumptions i) and ii) are hold, 
   then we can expect that nothing will substantially change when a narrow-band filter 
   transmitting only 
   frequencies in the viscinity of~$\omega_0$ will be installed between 
   photodetector and tank circuit. The LC circuit itself plays role of such a filter. 
   It means that before substituting the current $I$ from Eq.~(\ref{2}) to 
   (\ref{3}) one can omit all harmonics in it with the exception of~$I_\omega$. 
   Substituting~(\ref{7}) to~(\ref{2}), we obtain:
   \begin{equation}
   I_\omega= \cos(\omega t)\left(SA-{3S_3A^3\over 4}\right).
   \label{9}
   \end{equation}
    Then substituting to (\ref{3}) $I_\omega$ obtained by (\ref{9}), we get
   \begin{equation}
   A\sin(\omega t)\left({1\over \omega L}-\omega C\right)=
   A\cos(\omega t)\left({1\over \omega ^2\tau L}-S +{3\over 4}S_3A^2\right).
   \end{equation}
   To satisfy this equation for any~$t$ with nonzero~$A$, both parenthesized differences 
   must vanish. That can be written as two separate quadratic equations for $\omega$
   $$	{1\over\omega L}=\omega C $$
   and $A$
   $$	{3\over4}S_3A^2=S-{1\over\omega^2\tau L},$$
   which have their respective solutions
   \begin{equation}
   \omega={1\over \sqrt{LC}}
   \label{11}
   \end{equation}
   and
   \begin{equation}
   A=2\sqrt{S-RC/L\over 3S_3}.
   \label{12}
   \end{equation}
   Equation (\ref{11}) shows that assumptions i) and ii) are consistent
   and~(\ref{12}) illustrates critical behaviour of the oscillations 
   amplitude with excitation threshold exactly defined by relation~(\ref{5}).
   
   Due to the fact that $S$ and $S_3$ are both proportional to the light 
   intensity $P$ one can see that the sensitivity of the described oscillations 
   amplitude to the small light intesity fluctuations~$dA/dP$ diverges at the 
   excitation threshold. This allows one to use near-critically 
   excited oscillator as a sensitive photodetector.
    
   \section{Experiment}
   
   In experiment we used cylindrical coil 30 mm in length and 25 mm in diameter 
   comprising of two layers of 0.9 mm thick silver-plated copper wire. The tank 
   circuit was formed by this coil together with the sum of input capacitance of Pockels
   cell, output capacitance of photocell, internal coil capacitance and other
   parasitic capacitances. 
   We used linearly polarized 0.5 W argon laser as a light source, Pockels cell 
   combined with a Glan-Taylor prism ML-103 as an electro-optical modulator and 
   vacuum photoelectric cell F-22 with bulk photocathode as a photodetector. For 
   the sake of better linearity we installed a matte scatterer in front of the photocell 
   to uniformly illuminate the whole photocathode surface. The RF generation was 
   observed by means of the auxiliary coil L' (see Fig.~\ref{fig2}) 
   located 5~mm apart from the tank circuit. Amplitude of the RF voltage 
   observed on an 100~$\Omega$ load resistor reached $\sim 3$~V.
   
   As it was already mentioned, if the generation threshold is barely exceeded, 
   the amplitude of oscillations in the system strongly depends 
   on light intensity (see Eq.~(\ref{12})). This fact allows one to use 
   weakly excited generator as an ``active" photodetector. 
   In our experimental setup it was easy to obtain the regime when 5\% fluctuations
   of the laser beam intensity produced 1~V modulation of the output RF voltage on 
   100 $\Omega$ load resistor. This is two orders of magnitude greater than voltage
   observed 
   from the same intensity fluctuations 
   with the same photocell
   on the same 100 $\Omega$ load resistor 
   connected in conventional circuit (that is without coils and Pockels cell).
   
   \section{Conclusion}
   
   The paper describes an optical analogue of the electro-vacuum triode, in which 
the electron flux is replaced by a photon flux, and the grid is replaced by an 
electro-optical modulator. The properties of the proposed amplifying element are 
considered. A scheme of an RF generator on an optical triode is proposed and its 
quantitative analysis is given. The possibility of using an optical triode generator 
as an active photodetector is demonstrated. According to the authors, the described 
RF generator on an optical triode can serve to demonstrate the principles of 
designing amplifying elements and simple radio circuits based on them in school 
and student audiences.

\end{document}